# Ultraviolet radiation impact on the efficiency of commercial crystalline silicon-based photovoltaics: A theoretical thermal-electrical study in realistic device architectures

GEORGE PERRAKIS,[1,2,*] ANNA C. TASOLAMPROU,[1] GEORGE KENANAKIS,[1] ELEFTHERIOS N. ECONOMOU,[1,3] STELIOS TZORTZAKIS,[1,2,4] MARIA KAFESAKI[1,2]

[1]Institute of Electronic Structure and Laser (IESL), Foundation for Research and Technology-Hellas (FORTH), 70013 Heraklion, Crete, Greece
[2]Department of Materials Science and Technology, University of Crete, 70013 Heraklion, Crete, Greece
[3]Department of Physics, University of Crete, 70013 Heraklion, Crete, Greece
[4]Science Program, Texas A&M University at Qatar, P.O. Box 23874, Doha, Qatar
*Corresponding author: gperrakis@iesl.forth.gr

**We investigate and evaluate the contribution of the ultraviolet radiation spectrum on the temperature and efficiency of commercial crystalline silicon-based photovoltaics (PVs) that operate outdoors. The investigation is performed by employing a comprehensive thermal-electrical modeling approach which takes into account all the major processes affected by the temperature variation in the photovoltaic devices. We show that effectively reflecting the ultraviolet radiation (i.e. up to a certain wavelength) results in a reduction of the overall operation temperature and enhancement of the PV cell's efficiency. In addition, blocking the high energy ultraviolet photons prolongs the life time of the PV and its performance on the long term.**

Photovoltaics (PVs) are currently the fastest-growing solar technology [1]. Still, to be economically viable they need to operate consistently at outdoor conditions and to be reliable for at least 25 years. The deterioration of performance and the degradation factors have been shown to be directly linked to the wavelength and intensity of incident light, as well as the overall system temperature. For instance, elevated operating temperatures arising by the increased heating in a PV module decline its performance due to the increased carrier recombination and the subsequent voltage reduction [2] and eventually the reduction of the output electrical power. The operating temperature is directly dependent, among others, on the wavelength and the intensity of incident light. More specifically, the excess of energy of incident photons relative to the semiconductor's bandgap energy has been identified as the main factor for increasing heat in a solar cell [3]. This excess of energy cannot be exploited and finally dissipates into heat (often called thermalization losses), increasing significantly the temperature of the device and thus reducing its efficiency.

Moreover, PV's degradation over the years and its aging rate need to be treated more effectively for the long-term power stability of the device. A commercial PV's degradation is mainly attributed to the degradation of the eminent polymer encapsulant ethylene-vinyl-acetate (EVA) copolymer, employed as adhesion layer/layers between the cells. The photochemical processes caused by light, such as photodegradation, lead to the alteration of the primary structure of the polymer, due to breaking of the chemical bonds in its main chain, initiating unwanted reactions [4]. Such photochemical processes is a major material-degradation factor resulting even to reduced transmission from the EVA (yellowing) and thus harming PV's performance significantly [5]. As with most photochemical processes, the reaction rates depend on the wavelength, intensity of incident light, as well as the temperature.

Ultraviolet (UV) radiation has been identified as the most critical factor for the degradation of photovoltaics [4–6]. High energy UV photons (0.28–0.4 μm) can break chemical bonds in the main chain of the EVA-polymer as well as cause damage to the front surface of the silicon layer (i.e. defects, acting as recombination traps). As such, UV-originated cell and encapsulant degradation comprise the highest fraction (up to an even ~70%) of the 25-year power degradation [6] of PVs (assuming a standard 25-year warranty of less than 20% power loss). Moreover, UV light is associated with a high fraction of thermalization losses in PVs, leading to higher device temperatures and thus to negative impact on the PV efficiency.

Therefore, several approaches are used for blocking UV light. The most common approach is employing UV absorbers which are either photoactive chemicals (having though finite lifetimes due to photothermal oxidative degradation), or low iron (Fe) glasses [7] doped with cerium (Ce) (which are subject to oxidation, leading to absorption also of other beneficial parts of the spectrum and thus to PV performance reduction [4]). A more direct solution, by Kempe et al. [7] and Li et al. [8], is the utilization of antireflection coatings to reflect UV light. However, reflecting UV light, leads also to reduced light transmission inside the cell, resulting to less photocurrent and thus to negative impact on the PV efficiency.

Given the above, it is very important to examine and evaluate the currently unexplored total impact of the UV spectrum (wavelengths from ~0.28-0.4 μm) on the efficiency of a solar cell considering all UV associated and competing effects, i.e. the increase of the photocurrent on one hand and the increase of the temperature on the other hand (due to the high thermalization losses in the cell and the high parasitic absorption from EVA encapsulant that is located on top of the cell).

In this respect, in the present study, we consider realistic commercial crystalline silicon PVs (dominant in the market of solar cell technology [9]) and evaluate in detail the total impact of the UV radiation on the PV efficiency. For this evaluation we employ a comprehensive thermal-electrical co-model (described in detail in Perrakis et al. [10]) which calculates the solar cell steady state temperature (for given incident power, materials and weather conditions) as well as its efficiency as a function of temperature taking into account all the major processes affected by the temperature variation in a commercial PV device. These processes include the material-dependent radiative and non-radiative recombination of electron-hole pairs (such as the temperature-dependent nonradiative Auger recombination process), which have been identified as the major cause for the voltage decline and the subsequent efficiency decrease of PVs operating at elevated temperatures [10].

The model-PV system employed in the present study is a state-of-the-art silicon-based PV module [8,11] (see Fig. 1a), where the active layer (within the cell) is of crystalline silicon, basically a p-n homojunction diode (silicon bandgap ~1.107μm). The cell is placed in-between two EVA layers while a top glass layer is used to protect the cell and offer more stability. (The system is described in more detail in Fig. 1a, while the material parameters for its different materials are obtained from Ref. [10]). In this system we explore the impact of the UV radiation by gradually reflecting it (by 100%), starting from a wavelength equal to 0.28 μm - where the highest thermalization losses occur, up to a given wavelength $\lambda_r$, and calculating the output electrical power or efficiency with respect to the operating temperature ($T$) at typical outdoor conditions. (As a positive "side effect", directly reflecting UV radiation instead of utilizing UV absorbers results to a reduced overall system temperature.)

More specifically, our modelling approach [10] integrates (i) full-wave electromagnetic simulations (using the commercially available software CST Microwave Studio), to calculate the absorptivity/emissivity of the PV at the thermal mid-IR wavelengths (4-33 μm) (with a 5° angular resolution), and a (ii) thermal and (iii) electrical part. The full-wave calculated emissivity data are imported into the thermal part of our analysis, which associated to the reflection of the incident UV radiation for the system of (a), for a reflection wavelength range from 0.28 μm to λr. For all cases the ambient temperature is equal to 298 K. To mimic typical outdoor conditions, we assume an irradiance level (Irrl) of 40% (of the "AM 1.5G" standard sunlight spectrum [12]) and a combined nonradiative heat transfer coefficient, hc, equal to 20 W/m2/K (black lines), Irrl=100%, hc=20 W/m2/K (blue lines), and Irrl =100%, hc=10.6 W/m2/K (red lines). The green dashed line in (c) indicates the EVA absorption. The two black/red dashed vertical lines correspond to two different λr of 0.363/0.37 μm and 0.375/0.393 μm where we observe the maximum efficiency improvement and the limiting point where the efficiency remains unharmed for the conditions of the black/red curve case. The orange and blue filled areas in (b) and (c) correspond to the normalized "AM1.5G" standard sunlight spectral irradiance and photon flux respectively.

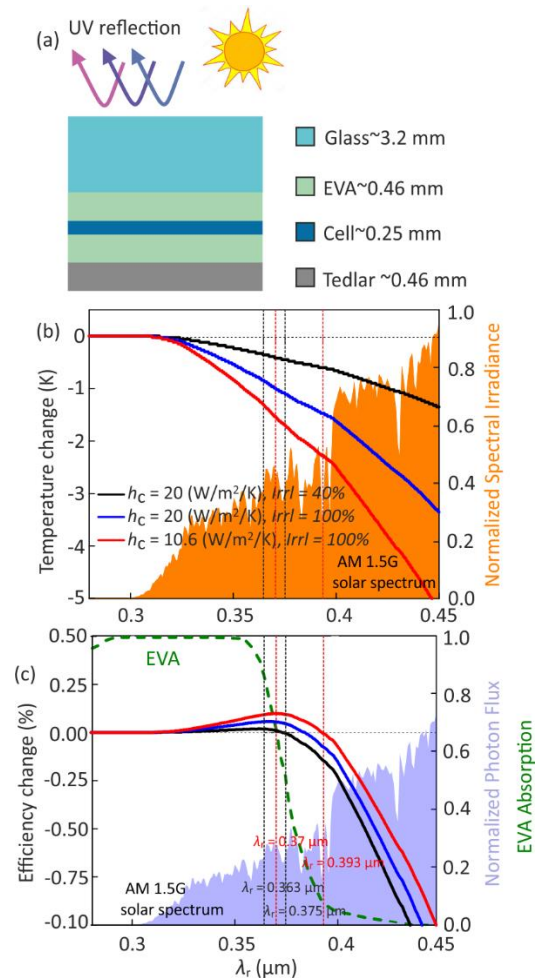

Fig. 1. (a) Schematic of the silicon-based PV module investigated in this work, showing the different layers along with their thickness. The material parameters and the absorptivity/emissivity data of the PV module in the optical spectral range are the same as in Ref. [10]. (b) PV temperature and (c) efficiency change

utilizes the passive radiative cooling approach that was firstly proposed by Fan [13] to calculate the steady state temperature of the device. The approach balances the power "into" and "out of" the device. The power "into" the device is the power absorbed from the solar emission, noted as $P_{sun}$, which is directly affected by any elimination of UV part, and the power absorbed from the atmosphere, $P_{atm}(T_{amb})$, where $T_{amb}$ is the ambient temperature. The power "out of" the device is connected with the thermal radiation, $P_{rad,PV}$, the non-thermal radiation, $P_{rad,cell}$, which corresponds to the radiation emitted through electron-hole recombination [14], the nonradiative heat transfer ($P_{cond+conv}=h_c(T-T_{amb})$, $h_c$ is the nonradiative heat transfer coefficient) and the output electrical power ($P_{ele,max}$). The above are summarized in the following relation:

$$P_{net,cool}(V,T) = P_{rad,PV}(T) - P_{atm}(T_{amb}) + P_{cond+conv}(T_{amb},T) - P_{sun} + P_{ele,max}(V,T) + P_{rad,cell}(V,T) \quad (1)$$

Since in our case the operating temperature of the device is greater than the ambient temperature, $T_{amb}$, the nonradiative heat transfer (due to the convection and the conduction taking place within the device and its interface with the environment) is a heat dissipation out of the device, offering an additional cooling effect besides radiative cooling. The electrical part utilizes the Detailed Balance Principle described by Shockley and Queisser [15] to calculate the current ($J$) - voltage ($V$) characteristics of the cell for a given temperature (and through them the power $P_{ele,max}$, plugged-in also in Equation (1)). In applying this principle, besides the losses due to radiative electron-hole recombination, we additionally take into consideration the fundamental temperature dependent non-radiative-Auger recombination loss [16]. In this respect, the efficiency, $\eta$, of a commercial crystalline silicon PV which operates at its maximum power (mp) point is given by

$$\eta = P_{ele,max} / P_{inc} = \max(-J \cdot V) / P_{inc} = J_{mp} \cdot V_{mp} / P_{inc} \quad (2)$$

where $P_{inc}$ is the incident power. The efficiency is self-consistently determined as we combine the electrical and the thermal part and solve the steady-state problem, i.e., when the power balance equals to zero [($P_{net,cool}(V_{mp},T)=0$ see Equation (1)] [10,13].

The above described approach (detailed in Ref. [10]), which allows to calculate the impact of any part of the electromagnetic spectrum on the PV efficiency, is employed below for the examination of the UV reflection impact on the efficiency and operating temperature of the realistic PV module shown in Fig. 1a operating at outdoor conditions.

Figures 1b and 1c depict the impact on the PV temperature change (Fig. 1b) and on the PV efficiency (Fig. 1c) of totally reflecting the solar energy from 0.28 μm to a parameter wavelength $\lambda_r$, as $\lambda_r$ varies from 0.28 μm to 0.45 μm, i.e. covers all the emitted by the sun UV radiation. For the temperature and efficiency calculations we assumed that the PV is operating at outdoor conditions, with $T_{amb}$=298 K. To capture the effect of the variant environmental conditions we assume three different cases, (i) an irradiance level (Irrl) of 40% (of the "AM 1.5G" standard sunlight spectrum [12]) and a combined conduction-convection nonradiative heat transfer coefficient, $h_c$, equal to 20 W/m$^2$/K (black curves), a value corresponding to strong wind climates, (ii) Irrl =100%, $h_c$=20 W/m$^2$/K (blue curves), and (iii) Irrl =100%, $h_c$=10.6 W/m$^2$/K (red curves), i.e. weak wind climates. The orange and blue areas in Fig. 1b and Fig. 1c respectively correspond to the normalized "AM1.5G" standard sunlight spectral irradiance and photon flux respectively and the green dashed curve in Fig. 1c indicates the EVA absorption.

As seen in Fig. 1b, reflecting incident radiation leads always to a temperature reduction (compared to the primary PV, i.e. without UV reflection), as is expected. Interestingly though, we found that the reflection of the UV radiation up to a certain wavelength may lead to an increase (up to ~0.1%) rather than a decrease of the PV efficiency, despite the reduction of potential carriers. This is clearly seen in Fig. 1c and, e.g., for 0.28–0.39 μm where the efficiency change obtains positive values in the red curve case. In other words, the negative effects of high EVA absorption (shown by the green dashed line in Fig. 1c) in this regime and the thermalization losses seem to overcompensate the positive effect of the additional potential carriers generated by the UV (see blue area in Fig. 1c).

Moreover, the impact of reflecting certain UV wavelengths on the device's temperature and efficiency varies for each of the different cases due to the alteration of the environmental conditions that affect the power-temperature relation and thus the steady-state operating temperature [10]. Climates with lower wind speeds, e.g. $h_c$ <13 W/m$^2$/K, are expected to allow higher cut-off wavelength $\lambda_r$ and hence higher temperature reduction. For instance, assuming a wider reflection wavelength range, with $\lambda_r$ =0.393 μm (see right vertical red line in Fig. 1b and Fig. 1c), the PV could operate at an up to ~2.3 K lower temperature compared to $\lambda_r$ =0.37 μm but its performance is not sacrificed only for the $h_c$=10.6 W/m$^2$/K, Irrl=100% case (where the efficiency change remains positive as seen in Fig. 1b - red curve). In implementing a practical approach though to cut parts of the UV spectrum, given that $\lambda_r$ can be specified only during the manufacturing procedure, it is essential to calculate a $\lambda_r$ which will be robust in respect to the variant environmental conditions as well as the various characteristics of commercial PVs.

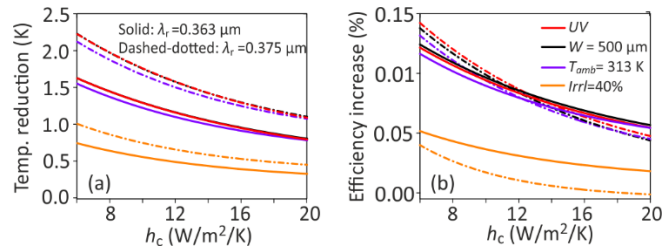

Fig. 2. (a) PV temperature reduction and (b) efficiency increase associated to the reflection of the incident UV radiation for a wavelength range from 0.28 μm up to λr =0.363 μm (solid lines) and λr =0.375 μm (dashed-dotted lines) for ambient temperature Tamb =298 K, with respect to the nonradiative heat transfer coefficient, hc, for the system of Fig. 1a. The figures show the impact of the UV reflection for an irradiance level (Irrl) 100% (UV – red lines), and a much lower one (Irrl =40% – orange lines), and for different PV characteristics, like higher silicon thickness (W =500 μm – black lines), Tamb =313K (purple lines).

Next we calculate such a constant/fixed cut-off reflection wavelength, $\lambda_r$; it is calculated for $T_{amb}$=298 K, $h_c$ =20 W/m$^2$/K, and a lower irradiance level (40% of the "AM 1.5G" standard sunlight spectrum [12]), which is the worst-case scenario studied (black curves of Fig. 1). Calculating $\lambda_r$ by requiring *maximum temperature reduction* (without harming efficiency) we obtain $\lambda_r$ =0.375 μm; requiring the *maximum possible efficiency increase* (for the above-mentioned environmental conditions) the resulting cut-off reflection wavelength is $\lambda_r$ =0.363 μm.

Figure 2 presents the impact of the UV reflection on temperature (Fig. 2a) and efficiency (Fig. 2b) for a PV operating outdoors for different, typical environmental conditions, i.e. as a function of the combined conduction-convection coefficient. The UV reflection is total reflection (i.e. 100%) in the wavelength range from 0.28 μm up to the two cut-off reflection wavelengths $\lambda_r$ specified above (solid curves correspond to $\lambda_r$ =0.363 μm, assumed for maximum efficiency, and dashed-dotted curves correspond to $\lambda_r$ =0.375 μm, assumed for maximum temperature reduction). Additionally, in Fig. 2 we examine the impact of the UV reflection on the PV temperature and efficiency for thicker silicon layer (W =500 μm – black), higher $T_{amb}$ (313 K - purple), and an irradiance level 100% (Irrl =100% – red) and a much lower one (Irrl =40% – orange). As seen in Fig. 2, reflecting UV radiation up to $\lambda_r$ =0.375 μm leads to an efficiency increase by up to ~0.15% (dashed-dotted red in Fig 2b). Additionally, the temperature reduction compared to the primary PV, i.e. without UV reflection, can reach values up to ~2.2 K (dashed-dotted red in Fig 2a). Assuming a narrower reflection wavelength range ($\lambda_r$ =0.363 μm) results to a slightly higher efficiency (compared to the $\lambda_r$ =0.375 μm case - see solid versus dashed-dotted lines in Fig 2b) for more windy climates ($h_c$ >13 W/m$^2$/K), but with slightly (up to ~0.5 K) higher operating temperature (see solid versus dashed-dotted curves in Fig 2a). Moreover, as seen in Fig. 2, the results are robust even for different PV or environment characteristics met in commercial PVs operating outdoors. Taking into account all cases, the PV operating temperature can be reduced by up to ~2.2 K (see Fig. 2a) due to the UV reflection, without decreasing the efficiency (see Fig. 2b). An even higher impact from UV reflection can be expected in PVs with high electrical-power - temperature coefficients, as well as in top contact solar cells, where there is additional parasitic absorption from the metallic top contacts (thus higher room for heat elimination), in concentrated systems, and in PVs with lower than unity internal quantum efficiencies [17] (collected carriers - absorbed photons ratio) in UV.

To demonstrate the UV-reflection impact with realistic structures we apply the theory discussed above in the case of the PV device of Fig. 1(a) covered by a realistic UV reflector, that is a one-dimensional (1D) photonic crystal (PC) – see Fig. 3(a). The proposed 1D PC consists of 45 alternating Si$_3$N$_4$ (ε∼4) – MgF$_2$ (ε∼1.82) thin layers and was designed to effectively reflect part of the UV spectrum (i.e. reflects as close as possible to the previously presented 0.28-$\lambda_r$ wavelengths) – see Fig. 3(b) (We note that in our study, we assume the thermal power radiated by the PV is not affected by the top thin 1D PC.).

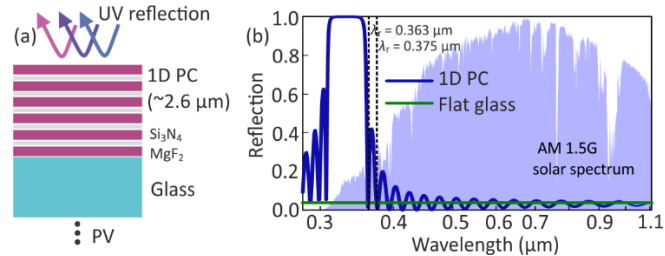

**Fig. 3.** (a) Illustration of a 1D photonic crystal (consisting of alternate Si$_3$N$_4$ – MgF$_2$ thin layers of 15 – 100 nm thickness respectively – total thickness ~2.6 μm) placed on top of the PV. (b) Reflectivity spectra of the 1D photonic crystal (blue line) in comparison with the reflectivity of flat glass (i.e. top layer in Fig. 1(a) – green line). The two black dashed vertical lines correspond to two different $\lambda_r$ of 0.363/0.375 μm discussed in connection with Fig. 2.

As can be seen in Fig. 3(b) the 1D PC not only reflects the high-energy photons (i.e. at lower wavelengths) but also increases PV's top surface transparency, compared to the flat glass, at the higher wavelengths. As a result, by implementing the 1D PC at the PV device of Fig. 1(a) and assuming $T_{amb}$=298 K, Irrl=100%, $h_c$=20 W/m$^2$/K, the PV efficiency increased by ~0.19% while the temperature decreased by 1 K. These results confirm that the utilization of suitable antireflective coatings, to reflect effectively UV, can lead to PVs' temperature reduction and lifetime increase without harming their efficiency or even enhancing this efficiency.

In conclusion, we examined the role of the ultraviolet spectrum on the efficiency of commercial crystalline silicon-based photovoltaics operating outdoors. Doing so, we observed that by reflecting ultraviolet radiation in the range 0.28 to 0.375 μm we were able to reduce the PV operating temperature by more than 2 K and increase considerably the system lifetime. The optimum reflection wavelength range for temperature reduction was found to be from 0.28 μm to 0.375 μm. Moreover, the proper reflection of the UV radiation and the associated operating temperature decrease can lead under certain conditions even to increase of the PV efficiency despite the "reflection" of potential carriers. Therefore, implementing a photonic approach to effectively reflect UV radiation can provide an effective alternative to the existing costly techniques for screening UV, increasing thus both the efficiency and the life time of the solar cells.

**References**

1. "Global Overview," https://www.ren21.net/gsr-2019/chapters/chapter_01/chapter_01/.
2. S. Dubey et al., Energy Procedia **33**, 311–321 (2013).
3. O. Dupré et al., Sol. Energy Mater. Sol. Cells **140**, 92–100 (2015).
4. M. C. C. de Oliveira et al., Renew. Sustain. Energy Rev. **81**, 2299–2317 (2018).
5. A. Ndiaye et al., Sol. Energy **96**, 140–151 (2013).
6. D. C. Jordan et al., Prog. Photovoltaics Res. Appl. **21**(1), 12–29 (2013).
7. Kempe M.D. et al., National Renewable Energy Laboratory, Conference Paper, NREL, Available from: ⟨http://www.nrel.gov/docs/fy09osti/44936.pdf⟩, (2009).
8. W. Li et al., ACS Photonics **4**(4), 774–782 (2017).
9. T. Saga, NPG Asia Mater. **2**(3), 96–102 (2010).
10. https://arxiv.org/abs/1912.12154.
11. D. D. Smith et al., in *2013 IEEE 39th Photovoltaic Specialists Conference (PVSC)* (IEEE), pp. 0908–0913 (2013).
12. "Solar Spectral Irradiance: Air Mass 1.5," https://rredc.nrel.gov/solar/spectra/am1.5/.
13. E. Rephaeli et al., Nano Lett. **13**(4), 1457–1461 (2013).
14. P. Wurfel, J. Phys. C Solid State Phys. **15**(18), 3967–3985 (1982).
15. W. Shockley and H. J. Queisser, J. Appl. Phys. **32**(3), 510–519 (1961).
16. M. A. Green, IEEE Trans. Electron Devices **31**(5), 671–678 (1984).
17. W. J. Yang et al., Sol. Energy **82**(2), 106–110 (2008).